\documentclass[%
 reprint,
superscriptaddress,
nofootinbib,
 amsmath,amssymb,
 aps,
]{revtex4-2}

\usepackage{graphicx}
\usepackage{dcolumn}
\usepackage{bm}
\usepackage{hyperref}
\usepackage{float}

\begin{document}

\title{Quantum computing models for artificial neural networks}

\author{Stefano Mangini}
\email{stefano.mangini01@universitadipavia.it}
\affiliation{Dipartimento di Fisica, Università di Pavia, Via Bassi 6, I-27100, Pavia, Italy}
\affiliation{INFN Sezione di Pavia, Via Bassi 6, I-27100, Pavia, Italy}

\author{Francesco Tacchino}
\affiliation{IBM Quantum, IBM Research - Zurich, Säumerstrasse 4, CH-8803 Rüschlikon, Switzerland}

\author{Dario Gerace}
\affiliation{Dipartimento di Fisica, Università di Pavia, Via Bassi 6, I-27100, Pavia, Italy}

\author{Daniele Bajoni}
\affiliation{Dipartimento di Ingegneria Industriale e dell’Informazione, Università di Pavia, Via Ferrata 1, 27100, Pavia, Italy}

\author{Chiara Macchiavello}
\affiliation{Dipartimento di Fisica, Università di Pavia, Via Bassi 6, I-27100, Pavia, Italy}
\affiliation{INFN Sezione di Pavia, Via Bassi 6, I-27100, Pavia, Italy}
\affiliation{CNR-INO -  Largo E. Fermi 6, I-50125, Firenze, Italy}

\date{\today}

\begin{abstract}
Neural networks are computing models that have been leading progress in Machine Learning (ML) and Artificial Intelligence (AI) applications. In parallel, the first small scale quantum computing devices have become available in recent years, paving the way for the development of a new paradigm in information processing. Here we give an overview of the most recent proposals aimed at bringing together these ongoing revolutions, and particularly at implementing the key functionalities of artificial neural networks on quantum architectures. We highlight the exciting perspectives in this context, and discuss the potential role of near term quantum hardware in the quest for quantum machine learning advantage.
\end{abstract}

\maketitle

\section{Introduction}
Artificial Intelligence (AI) broadly refers to a wide set of algorithms that have shown in the last decade impressive and sometimes surprising successes in the analysis of large datasets \cite{Goodfellow2016DeepL}. These algorithms are often based on a computational model called Neural Network (NN), which belongs to the family of differential programming techniques. 

\begin{figure*}[htbp]
    \centering
    \includegraphics[scale=0.25]{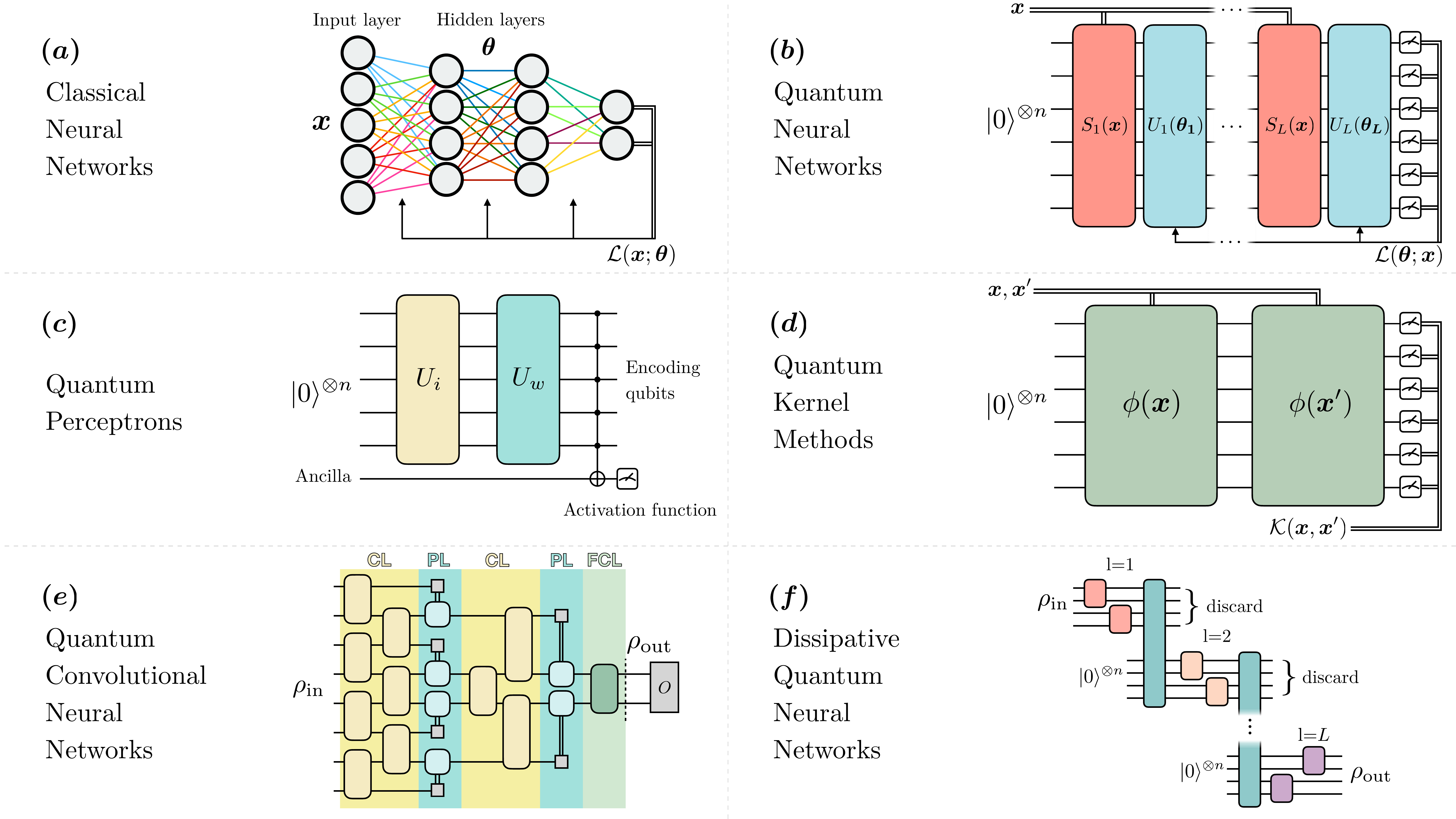}
    \caption{{\bf Schematic representation of classical and quantum learning models}. {\bf(a)} Classical feedforward NNs process information through consecutive layers of nodes, or \textit{neurons}. {\bf(b)} A quantum neural network in the form of a variational quantum circuit: the data \textit{re-uploading} procedure ($S_i(\bm{x})$) is explicitly illustrated, and non-parametrized gates ($W_i$) of Eq.~\ref{eq:qnn1} are incorporated inside the variational unitaries ($U_i(\bm{\theta}_i)$). {\bf(c)} A possible model for a quantum perceptron \cite{Tacchino_2019}, which uses operations $U_i$ and $U_w$ to load input data and perform multiplication by weights, and a final measurement is used to emulate the neuron activation function. {\bf(d)} Quantum Kernel Methods map input data to quantum states \cite{Havlicek2019SVM}, and then evaluate their inner product $\langle \phi{(\bm{x})}|\phi{(\bm{x}')}\rangle$ to build the Kernel matrix $\mathcal{K}(\bm{x},\bm{x}')$. {\bf(e)} A Quantum Convolutional NN \cite{Lukin2019Convolutional} consists of a repeated application of Convolutional layers (CL) and Pooling layers (PL), ultimately followed by a fully-connected layer (FCL), i.e., a general unitary on all remaining qubits applied before a measurement (or else, $O$). {\bf(f)} In a Dissipative QNN \cite{Bondarenko2020TrainingDeep}, qubits in one layer are coupled to ancillary qubits from the following layer in order to load the result of the computation, while previous layers are discarded (i.e., \textit{dissipated}). }
    \label{fig:QNNs}
\end{figure*}

The simplest realization of a NN is in a ``feedforward'' configuration, mathematically described as a concatenated application of affine transformations and element wise nonlinearities, $f_j ({\cdot}) = \sigma_j(\bm{w}_{j}{\cdot} + b_j)$, where $\sigma_j$ is a nonlinear \textit{activation function}
, $\bm{w}_j$ is a real matrix containing the so-called \textit{weights} to be optimized, and $b_j$ is a \textit{bias} vector, which is also to be optimized. 
The action of the NN is defined through repeated applications of this function to an input vector, $x$, leading to an output $\hat{y} = f_{L}\circ f_{L-1} \circ \cdots \circ f_1(x)$, where $L$ denotes the number of layers in the NN architecture, as schematically shown in Fig.~\ref{fig:QNNs}(a). 
The success of NNs as a computational model is primarily due to their trainability \cite{Hinton2006BM}, i.e.\ the fact that the entries of the weight matrices can be adjusted to learn a given target task. The most common use case is \textit{supervised learning}, where the algorithm is asked to reproduce the associations between some inputs $x_i$ and the desired correct outputs (or label) $y_i$. If properly trained, the NN should also be able to predict well on new data, i.e., data that were not used during training. This crucial property is called \textit{generalization}, and it represents the key aspect of learning algorithms, which ultimately distinguish them from standard fitting techniques. Generalization encapsulates the idea that the learning models discover hidden patterns and relations in the data, instead of applying pre-programmed procedures \cite{LeCun2015DeepL, Hastie2009Statistical, Goodfellow2016DeepL}.


The size of the data and features elaborated by classical NNs, and the number of parameters defining their structure, have been steadily increasing over the last years. 
These advancements have come at the expense of a dramatic rise in the demand for energy and computational power \cite{Edler_Global_Electricity_Usage} from the field of AI. In these regards, quantum computers may offer a viable route forward to continue such growth in the dimension of the treatable problems and, in parallel, enable completely new functionalities. In fact, quantum information is elaborated by exploiting superpositions in vector spaces (the so called Hilbert spaces) \cite{Nielsen_Chuang}, and quantum processors are planned to soon surpass the computational capabilities of classical supercomputers. Moreover, genuinely quantum correlations represent by themselves a resource to be considered in building new forms of ML applications.

With respect to classical computational paradigms, quantum computing comes with its own peculiarities. One of these is that quantum computation is intrinsically {\it linear}. While this property could ensure, e.g., that an increase in computational power does not come with a parallel increase in energy costs, it also means that implementing the nonlinear functions forming the backbone of NNs is a non-trivial task. Also, information cannot be copied with arbitrary precision in quantum algorithms, meaning that tasks such as repeatedly addressing a variable are impossible when the latter is represented by a quantum state. As a consequence, classical NN algorithms cannot be simply ported to quantum platforms.

In this article, we briefly review the main directions that have been taken through the years to develop artificial NNs for quantum computing environments, the strategies used to overcome the possible difficulties, and the future prospects for the field.

\section{Quantum Neural Networks} 

In gate-based quantum computers, such as those realized with superconducting qubits, ion traps, spins in semiconductors, or photonic circuits~\cite{tacchinoAQT2020}, a quantum computation is fully described by its quantum circuit representation, i.e., the arrangement of subsequent operations (gates) applied to the qubits in the quantum processing unit. These gates can be parametrized by continuous variables, like a Pauli-X rotation, $R_x(\theta) = e^{i\theta X/2}$. A quantum circuit that makes use of parametrized gates is known as a \textit{Parametrized Quantum Circuit} (PQC)~\cite{Benedetti2019Parametrized}. \textit{Variational Quantum Algorithms} (VQAs)~\cite{McClean2016VQAs, Peruzzo2014VQE, Cerezo2020VQAReview} are a large class of PQC-based protocols whose goal is to properly tune the parameters in order to solve a target task. For example, the task might be the minimization of the expectation value of an observable $\langle O \rangle = \langle \bm{0} |U(\bm{\theta})^\dagger O U(\bm{\theta}) |\bm{0}\rangle$, where $U(\bm{\theta})$ is the unitary operator representing the quantum circuit parametrized by angles $\bm{\theta}=(\theta_1,\theta_2, \hdots)$ and acting on a quantum register initialized in a reference state $|\bm{0}\rangle = |0\rangle^{\otimes n}$. VQAs are made by three main constituents: data encoding, variational \textit{ansatz}, and final measurements with a classical update of the parameters. The first stage accounts for loading input data (classical or quantum) into the quantum circuit using quantum operations that depend on the given input; the second stage consists of a variational circuit with a specific structure, called ansatz, which often consists of repeated applications of similar operations (\textit{layers}). Finally, a measurement on the qubits is performed to infer some relevant information about the system. Given the outcome, the parameters in the circuits are updated through a classical optimizer to minimize the cost function defining the problem. Such algorithms generally require limited resources to be executed on real quantum devices, thus they are expected to take full advantage of current noisy quantum hardware~\cite{Preskill2018NISQ, Bharti2021NISQ}.

The definitions above highlight many similarities between PQCs and classical NNs. In both models, information is processed through a sequence of parametrized \textit{layers}, which are iteratively updated using a classical optimizer. The key difference lies in the way information is processed. Quantum computers promise to possibly achieve some form of computational advantage, i.e., speedups or better performances~\cite{Biamonte2017QML}, due to inherently quantum effects not available in classical learning scenarios~\footnote{Earlier works on quantum generalizations of classical machine learning focused on the use of quantum computers to perform linear algebra exponentially faster compared to classical computers~\cite{Biamonte2017QML}. However, due to their unsuitability for NISQ devices and comparable performances with new classical \textit{quantum-inspired} methods~\cite{Aaronson2015Fine, Tang2019Dequantization, Arrazola2020Quantuminspired}, interest in this field is declining, with most of the research now redirected towards parametrized quantum circuits as quantum counterparts of classical learning models.}
. While many different definitions of Quantum Neural Networks (QNN) are possible, the most general and currently accepted one is that of a parametrized quantum circuit whose structure is directly inspired by classical neural networks. Thus, we formally define a QNN as a PQC whose variational ansatz contains multiple repetitions of self-similar layers of operations, that is:
\begin{eqnarray}
\label{eq:qnn1}
    U_{\text{QNN}}(\bm{\theta})  & = & \prod_{i=L}^1 U_i(\bm{\theta}_i)W_i \nonumber \\
    & = &  U_L(\bm{\theta}_L)W_L\cdots U_1(\bm{\theta}_1)W_1\, , 
\end{eqnarray}
where $U_i(\theta_i)$ are variational gates, $W_i$ are (typically entangling) fixed operations that do not depend on any parameter, and $L$ is the number of layers (see Fig.~\ref{fig:QNNs}(b) for a schematic picture). Depending on the problem under investigation and the connectivity of the quantum device (i.e., related to the native gate set available~\cite{tacchinoAQT2020}), various realizations exist for both the entangling structure and the parametrized gates~\cite{Cerezo2020VQAReview}. Note that this definition almost coincides with that of VQAs, and in fact the terms VQAs and QNNs are often used interchangeably.

While being rather general, a growing body of research has recently been highlighting that such a structure, in its naive realization, may be poorly \textit{expressive}, in the sense that it can only represent a few functions --- actually, mostly sine functions --- of the input data~\cite{Theis2020Expressivity, Schuld2020Encoding, Perez2020Reuploading, Mitarai2018Learning, Ostaszewski2019CircuitLearning}. A milestone of classical learning theory, the \textit{Universal Approximation Theorem}~\cite{Hornik1991Approximator}, guarantees that classical NNs are powerful enough to approximate any function $f(\bm{x})$ of the input data $\bm{x}$. Interestingly, it was shown that a similar result also holds for QNNs, provided that input data are loaded into the quantum circuit multiple times throughout the computation. Intuitively, such data \textit{re-uploading} can be seen as a necessary feature to counterbalance the effect of the \textit{no-cloning theorem} of quantum mechanics \cite{Nielsen_Chuang}. In fact, while in classical NNs the output of a neuron is copied and transferred to every neuron in the following layer, this is impossible in the quantum regime, thus it is necessary to explicitly introduce some degree of classical redundancy. As a rule of thumb, the more often data are re-uploaded during the computation, the more \textit{expressive} the quantum learning model becomes, as it is able to represent higher order features of the data. Thus, for a more effective construction, each layer in Eq.~\ref{eq:qnn1} should be replaced with $U_iW_i \rightarrow S(\bm{x})U_i W_i$, where $S(\bm{x})$ denotes the data encoding procedure mapping input $\bm{x}$ to its corresponding quantum state.

This formulation still leaves a lot of freedom in choosing how the encoding and the variational circuit are implemented~\cite{LaRose2020Robust, Lloyd2020Embeddings}. However, as it happens in the classical domain, also different architectures for QNNs exist, and in the following, we give a brief overview of some of the ones that received more attention. 

\subsubsection{Quantum Perceptron Models}
Early works on quantum models for single artificial neurons (known as \textit{perceptrons}) were focused on how to reproduce the non-linearities of classical perceptrons with quantum systems, which, on the contrary, obey linear unitary evolutions~\cite{Schuld2014Quest, Schuld2015Simulating, Cao2017quantum, Torrontegui2019Perceptron}. In more recent proposals (e.g., Fig.~\ref{fig:QNNs}(c)), quantum perceptrons were shown to be capable of carrying out simple pattern recognition tasks for binary~\cite{Tacchino_2019} and gray scale~\cite{Mangini_2020} images, possibly employing variational unsampling techniques for training and to find adaptive quantum circuit implementations~\cite{Tacchino2020IEEE}. By connecting multiple quantum neurons, models for quantum artificial neural networks were also realized on proof-of-principle experiments on superconducting devices~\cite{Tacchino2020Network}.

\subsubsection{Kernel methods} 
Directly inspired by their classical counterparts, the main idea of \textit{kernel methods} is to map input data into a higher dimensional space, and leverage this transformation to perform classification tasks otherwise unfeasible in the original, low dimensional representation. Typically, any pair of data vectors, $\bm{x},\bm{x}' \in \mathcal{X}$, is mapped into quantum states, $|\phi(\bm{x})\rangle$, $|\phi(\bm{x}')\rangle$, by means of an encoding mapping $\phi: \mathcal{X} \rightarrow \mathcal{H}$, where $\mathcal{H}$ denotes the Hilbert space of the quantum register. The inner product $\langle \phi(\bm{x})|\phi(\bm{x}')\rangle$, evaluated in an exponentially large Hilbert space, defines a similarity measure that can be used for classification purposes. The matrix $\mathcal{K}_{ij} = \langle \phi(\bm{x}_i)|\phi(\bm{x}_j)\rangle$, constructed over the training dataset, is called \textit{kernel}, and can be used along with classical methods such as Support Vector Machines (SVMs) to carry out classification~\cite{Havlicek2019SVM, Schuld2019FeatureSpace} or regression tasks, as schematically displayed in Fig.~\ref{fig:QNNs}(d). A particularly promising aspect of such quantum SVMs is the possibility of constructing kernel functions that are hard to compute classically, thus potentially leading to quantum advantage in classification.

\subsubsection{Generative Models}
A generative model is a probabilistic algorithm whose goal is to reproduce an unknown probability distribution $p_\mathcal{X}$ over a space $\mathcal{X}$, given a training set, $\mathcal{T}=\{\bm{x}_i|\, \bm{x}_i \sim p_\mathcal{X},\, i=1,\hdots,M\}$. A common classical approach to this task leverages probabilistic NNs called Boltzmann Machines (BMs)~\cite{Ackley198BM, Hinton2006BM}, which are physically inspired methods resembling Ising models. Upon changing classical nodes with qubits, and substituting the energy function with a quantum Hamiltonian, it is possible to define Quantum Boltzmann Machines (QBMs)~\cite{Amin2018Boltzmann, Zoufal2020Variational,Kierefova2017Boltzmann}, dedicated to the generative learning of quantum and classical distributions. On a parallel side, a novel and surprisingly effective classical tool for generative modeling is represented by \textit{Generative Adversarial Networks} (GANs). Here, the burden of the generative process is offloaded onto two separate neural networks called \textit{generator} and \textit{discriminator}, respectively, which are programmed to compete against each other. Through a careful and iterative learning routine, the generator learns to produce high-quality new examples, eventually fooling the discriminator. A straightforward translation to the quantum domain led to the formulation of Quantum GANs (QGANs) models~\cite{Zoufal2019QGAN, Lloyd2018QGANs, Demers2018QGANs}, where the generator and discriminator consist of QNNs. Even if not inspired by any classical neural network model, it is also worth mentioning Born machines~\cite{Coyle2020Born}, where the probability of generating new data follows the inherently probabilistic nature of quantum mechanics represented by Born's rule.

\subsubsection{Quantum Convolutional Neural Networks}
As the name suggests, these models are inspired by the corresponding classical counterpart, nowadays ubiquitous in the field of image processing. In these networks, schematically shown in Fig.~\ref{fig:QNNs}(e), inputs are processed through a repeated series of so-called \textit{convolutional} and \textit{pooling} layers. The former applies a convolution of the input---i.e., a quasi-local transformation in a translationally invariant manner--- to extract some relevant information, the latter applies a compression of such information in a lower dimensional representation. After many iterations, the network outputs an informative and low dimensional enough representation that can be analyzed with standard techniques, such as fully connected feedforward NNs. Similarly, in Quantum Convolutional NNs~\cite{Grant2018Hierarchical,Lukin2019Convolutional} a quantum state first undergoes a convolution layer, consisting of a parametrized unitary operation acting on individual subsets of the qubits, and then a pooling procedure obtained by measuring some of the qubits. This procedure is repeated until only a few qubits are left, where all relevant information was stored. This method proved particularly useful for genuinely quantum tasks such as quantum phase recognition and quantum error correction. Moreover, the use of only  logarithmically many parameters with respect to the number of qubits is particularly important to achieve efficient training and effective implementations on near-term devices.

\subsubsection{Quantum Dissipative Neural Networks}
Dissipative QNNs, as proposed in~\cite{Bondarenko2020TrainingDeep}, consist of models where each qubit represents a node, and edges connecting qubits in different layers represent general unitary operations acting on them. The use of the term \textit{dissipative} is related to the progressive discarding of layers of qubits after a given layer has interacted with the following one (see, e.g., Fig.~\ref{fig:QNNs}(f) for a schematic illustration). This model can implement and learn general quantum transformations, carrying out a universal quantum computation. Endowed with a backpropagation-like training algorithm, it represents a possible route towards the realization of quantum analogues of deep neural networks for analyzing quantum data.

\section{Trainability of QNN}
\label{sec:Trainability}
For decades, AI methods remained merely academic tools, due to the lack of computational power and proper training routines, 
which made the use of learning algorithms practically unfeasible on available devices. This underlines a very pivotal problem of all learning models, i.e., the ability to efficiently train them. 

\subsubsection{Barren plateaus}
\label{subsec:Barren}
Unfortunately, also quantum machine learning models were shown to suffer from serious trainability issues, which are commonly referred to as \textit{barren plateaus}\cite{McClean2018Barren}. In particular, using a parametrization as in Eq.~\eqref{eq:qnn1}, and assuming that such architecture forms a so-called unitary 2-design~\footnote{Roughly speaking, it is sufficiently random that sampling over its distribution matches the Haar distribution --- a generalization of the uniform distribution in the space of unitary matrices --- up to the second moment.}
, it can be shown that the derivative of any cost function of the form $C(\bm{\theta}) = \langle\bm{0} |U_{\text{QNN}}(\bm{\theta})^\dagger O U_{\text{QNN}}(\bm{\theta})|\bm{0}\rangle$ with respect to any of its parameters $\theta_k$, yields $\langle\partial_k C \rangle = 0$, and $\text{Var}[\partial_k C] \approx 2^{-n}$, where $n$ is the number of qubits. This means that, on average, there will be no clear direction of optimization in the cost function landscape upon random initialization of the parameters in a large enough circuit, being it essentially flat everywhere, except in close proximity to the minima. Moreover, barren plateaus were shown to be strongly dependent on the nature of the cost function driving the optimization routine~\cite{Cerezo2020CostBarren}. In particular, while the use of a global cost function always hinders the trainability of the model (independently of the depth of the quantum circuit), a local cost function allows for an efficient trainability for circuit depths of order $O(\log n)$. Besides, also entanglement plays a major role in hindering the trainability of QNNs. In fact, it was shown that whenever the objective cost function is evaluated only on a small subset of available qubits (visible units), with the remaining ones being traced out (hidden units), a barren plateau in the optimization landscape may occur~\cite{Marrero2020Entanglement}. This arises from the presence of entanglement, which tends to spread information across the whole network instead of its single parts. Moreover, also the presence of hardware noise during the execution of the quantum circuit can itself be the reason for the emergence of noise-induced barren plateaus~\cite{Wang2020NoiseBarren}. 

It is worth noticing that most of these results stem from strong assumptions on the shape of the variational ansatz, for example, its ability to approximate an exact 2-design. While these assumptions are often well-matched by proposed quantum learning models, QNNs should be analyzed case-by-case to have precise statements about their trainability. As an example, dissipative QNNs were shown to suffer from barren plateaus~\cite{Sharma2020DissBarren} when global unitaries are used, while convolutional QNNs seem to be immune to the barren plateaus phenomena~\cite{Pesah2020ConvBarren}. In general terms, however, recent results indicate that the trainability of quantum models is closely related to their expressivity, as measured by their distance from being an exact 2-design. Highly expressive ansatzes generally exhibit flatter landscapes and will be harder to train~\cite{Holmes2021Connecting}. At last, it was recently argued that the training of any VQA is by itself an NP-hard problem~\cite{Bittel2021TrainingNPHard}, thus being intrinsically difficult to tackle. Despite their intrinsic theoretical value, these results should not represent a severe threat, as similar conclusions also hold in general for classical NNs, particularly in worst-case analysis, and did not prevent their successful application in many specific use cases.

\subsubsection{Avoiding plateaus} 
\label{subsec:AvoidingBarren}
Several strategies have been proposed to avoid, or at least mitigate, the emergence of barren plateaus. A possible idea is to initialize and train parameters in batches, in a procedure known as layerwise learning~\cite{Skolik2020Layerwise}. Similarly, another approach is to reduce the effective depth of the circuit by randomly initializing only a subset of the total parameters, with the others chosen such that the overall circuit implements the identity operation~\cite{Grant2019Initialization}. Introducing correlations between the parameters in multiple layers of the QNN, thus reducing the overall dimensionality of the parameter space~\cite{Volkoff2020Correlating}, has also been proposed. Along a different direction, leveraging classical recurrent NNs to find good parameter initialization heuristics, such that the network starts close to a minimum, has also proven effective~\cite{Verdon2019Learning}. Finally, as previously mentioned, it is worth reminding that the choice of a local cost function and of specific entanglement scaling between hidden and visible units is of key importance to ultimately avoid barren plateaus~\cite{Cerezo2020CostBarren}.

\subsubsection{Optimization routines}
\label{subsec:Optimization}
The most common training routine in classical NNs is the backpropagation algorithm~\cite{Goodfellow2016DeepL}, which combines the chain rule for derivatives and stored values from intermediate layers to efficiently compute gradients. However, such a strategy cannot be straightforwardly generalized to the quantum case, because intermediate values from a quantum computation can only be accessed via measurements, which disturb the relevant quantum states~\cite{Schuld2019Gradients}. For this reason, the typical strategy is to use classical optimizers leveraging gradient-based or gradient-free numerical methods. However, there exist scenarios where the quantum nature of the optimization can be taken into account, leading to strategies tailored to the quantum domain like the \textit{parameter shift rule}~\cite{Mitarai2018Learning,Schuld2019Gradients}, the Quantum Natural Gradient~\cite{Stokes2020Quantumnatural}, and closed update formulas~\cite{Ostaszewski2019CircuitLearning}. Finally, while a fully coherent quantum update of the parameters is certainly appealing~\cite{Verdon2018universal}, this remains out of reach for near-term hardware, which is limited in the number of available qubits and circuit depth~\cite{tacchinoAQT2020}.

\section{Quantum advantage}
\label{sec:Advantage}
Quantum computers are expected to become more powerful than their classical counterparts on many specific applications~\cite{Harrow2017Sumpremacy, Google2019Supremacy, Zhong2020PhotonAdvantage}, and particularly in sampling from complex probability distributions --- a feature particularly relevant for generative modeling applications~\cite{Sweke2020PACLearning, Du2020ExpressivePQC}. It is therefore natural to ask whether this hierarchy also applies to learning models. While a positive answer is widely believed to be true, clear indications on how to achieve such a quantum advantage are still under study, with only a few results in this direction.

\subsubsection{Power of Quantum Neural Networks}
In the previous sections, we used the term \textit{expressivity} quite loosely. In particular, we used it with two different meanings: the first to indicate the ability of QNNs to approximate any function of the inputs~\cite{Schuld2020Encoding}, the second to indicate their distance from being an exact 2-design, i.e.\ the ability to span uniformly the space of quantum states~\cite{Holmes2021Connecting}. Each definition conveys a particular facet of an intuitive sense of \textit{power}, highlighting the need for a unifying measure of such power yet to be discovered. Nevertheless, some initial results towards a thorough analysis of the capacity (or power) of quantum neural networks have recently been put forward~\cite{Wright2019Capacity, Abbas2020Power, Huang2020Power}.

Starting from the Vapnik–Chervonenkis (VC) dimension, which is a classical measure of \textit{richness} of a learning model related to the maximum number of independent classifications that the model can implement, a universal metric of capacity related to the maximum information that can be stored in the parameters of a learning model, be it classical or quantum, was defined \cite{Wright2019Capacity}. Based on this definition, it was claimed that quantum learning models have the same overall capacity as any classical model having the same number of weights. However, as argued in Ref.~\cite{Abbas2020Power}, measures of capacity like the VC-dimension can become vacuous and difficult --- if not completely impractical --- to be measured, and therefore cannot be used as a general faithful measure of the actual power of learning models. For this reason, the use of \textit{effective dimension} as a measure of power is proposed, motivated by information-theoretical standpoints. This measure can be linked to generalization bounds of the learning models, leading to the result that quantum neural networks can be more expressive and efficient during training than their classical counterparts, thanks to a more evenly spread spectrum of the Fisher Information. Some evidence was provided through the analysis of two particular QNNs instances: one having a trivial input data encoding, and the other using the encoding scheme proposed in~\cite{Havlicek2019SVM}, which is thought to be classically hard to simulate and indeed shows remarkable performances. Despite such specificity, this is one of the first results drawing a clear separation between quantum and classical neural networks, highlighting the great potentials of the former.

\subsubsection{The role of data} 
In the same spirit, but with different tools, the role of data in analyzing the performances of quantum machine learning algorithms was taken into account \cite{Huang2020Power}, by devising geometric tests aimed at clearly identify those scenarios where quantum models can achieve an advantage. In particular, focusing on kernel methods, a geometric measure based exclusively on the training dataset and the kernel function induced by the learning model, which is actually independent of the actual classification/function to be learned, was defined. If such quantity is similar for the classical and quantum learning case, then classical models are proved to perform similarly or better than quantum models. However, if this geometric distance is large, then there exists some dataset where the quantum model can outperform the classical one. Notably, one key result of the study is that poor performances of quantum models could be attributed precisely to the careless encoding of data into exponentially large quantum Hilbert spaces, spreading them too far apart from each other and resulting in the kernel matrix approximating the identity. To avoid this setback, authors propose the idea of \textit{projected quantum kernels}, which aim at reducing the effective dimension of the encoded data set while keeping the advantages of quantum correlations. On engineered data sets, this procedure was shown to outperform all tested classical models in prediction error. From a higher perspective, the most important handout of these results is that data themselves play a major role in learning models, both quantum and classical.

\subsubsection{Quantum data for quantum neural networks}
Some data set could be particularly suited for QNNs. An example is provided by those proposed in~\cite{Liu2020rigorous} and based on the discrete logarithm problem. Also, it is not hard to suppose that quantum machine learning will prove particularly useful when analyzing data that are inherently quantum, for example coming from quantum sensors, or obtained from quantum chemistry simulations of condensed matter physics. However, even if this sounds theoretically appealing, no clear separation has yet been found in this respect. A particularly relevant open problem concerns, for example, the identification and generation of interesting quantum dataset. Moreover, based on information-theoretic derivations in the quantum communication formalism, it was recently shown \cite{Preskill2021Informationtheoretic} that a classical learner with access to quantum data coming from quantum experiments performs similarly to quantum models with respect to \textit{sample complexity} and \textit{average} prediction error. Even if an exponential advantage is still attainable if the goal is to minimize the \textit{worst-case} prediction error, classical methods are again found to be surprisingly powerful when given access even to quantum data.

\subsubsection{Quantum-inspired methods}
When discussing the effectiveness of quantum neural networks, and in order to achieve a true quantum advantage, one should compare not only to previous quantum models but also to the best available classical alternatives. For this reason, and until fault-tolerant quantum computation is achieved, quantum-inspired procedures on classical computers could pave the way towards the discovery of effective solutions~\cite{Aaronson2015Fine, Tang2019Dequantization, Arrazola2020Quantuminspired}. One of the most prominent examples is represented by tensor networks~\cite{Montangero_book}, initially developed for the study of quantum many-body systems, which show remarkably good performances for the simulation of quantum computers in specific regimes~\cite{Stoudenmire2020Limits}. In fact, many proposals for quantum neural networks are inspired by or can be directly rephrased in the language of tensor networks~\cite{Lukin2019Convolutional, Grant2018Hierarchical}, so that libraries developed for those applications could be used to run some instances of these quantum algorithms. Fierce competition is therefore to be expected in the coming years between quantum processors and advanced classical methods, including the so called neural network quantum states~\cite{Carleo602,torlai_many-body_2018} for the study many-body physics.


\section{Outlooks}
Quantum machine learning techniques, and particularly quantum neural network models, hold promise to significantly increase our computational capabilities, unlocking whole new areas of investigation. 
The main theoretical and practical challenges currently encompass trainability, the identification and suitable treatment of appropriate data sets, and a systematic comparison with classical counterparts. Moreover, with steady progress on the hardware side, significant attention will be devoted in the near future towards experimental demonstrations of practical relevance. Research evolves rapidly, with seminal results highlighting prominent directions to explore in the quest for quantum computational advantage in AI applications, and new exciting results are expected in the years to come. \\

\section{Acknowledgments}
This research was partly supported by the Italian Ministry of Education, University and Research (MIUR) through the ``Dipartimenti di Eccellenza Program (2018-2022)'', Department of Physics, University of Pavia, and the PRIN-2017 project 2017P9FJBS ``INPhoPOL'', and by the EU H2020 QuantERA ERA-NET Cofund in Quantum Technologies project QuICHE.

\bibliography{bibliography}

\end{document}